\title{Application of Machine Learning in Rock Facies Classification with Physics-Motivated Feature Augmentation}
\title{Automatic Seismic Salt Interpretation with Deep Convolutional Neural Networks}
\author{
Yu Zeng$^1$$^\dagger$, Kebei Jiang$^2$, and Jie Chen$^3$ \\
\\
$^1$ yu.zeng@alumni.duke.edu \\
$^2$ kebei.jiang13@gmail.com   \\
$^3$ jie.ch.2000@gmail.com   \\
\\
$^{\dagger}$ Corresponding author \\
}

\documentclass[12pt]{article}
\usepackage{amsmath}
\usepackage{graphicx}
\usepackage{multicol}
\usepackage{cuted}
\usepackage{subfig}
\usepackage{booktabs}
\usepackage{natbib}
\usepackage{epsfig}           
\usepackage{graphics}         
\usepackage{fancyheadings}          
\usepackage{float}          
\usepackage{times}       
\usepackage{amsmath,amsfonts,amssymb,amscd,amsthm,xspace}         
\usepackage{algorithmic}         
\usepackage{algorithm2e}        
\usepackage{verbatim}        
\usepackage{appendix}         
\usepackage{hyperref}
\usepackage[left=2cm,
right=2cm,
top=3cm, 
bottom=3cm,
a4paper
]{geometry}
\usepackage{setspace}

\begin{document}
\maketitle

\begin{abstract}
One of the most crucial tasks in seismic reflection imaging is to identify the salt bodies with high precision. Traditionally, this is accomplished by visually picking the salt/sediment boundaries, which requires a great amount of manual work and may introduce systematic bias. With recent progress of deep learning algorithm and growing computational power, a great deal of efforts have been made to replace human effort with machine power in salt body interpretation. Currently, the method of Convolutional neural networks (CNN) is revolutionizing the computer vision field and has been a hot topic in the image analysis. In this paper, the benefits of CNN-based classification are demonstrated by using a state-of-art network structure U-Net, along with the residual learning framework ResNet, to delineate salt body with high precision. Network adjustments, including the Exponential Linear Units (ELU) activation function, the Lov\'{a}sz-Softmax loss function, and stratified $K$-fold cross-validation, have been deployed to further improve the prediction accuracy. The preliminary result using SEG Advanced Modeling (SEAM) data shows good agreement between the predicted salt body and manually interpreted salt body, especially in areas with weak reflections. This indicates the great potential of applying CNN for salt-related interpretations.



\end{abstract}

\section{Introduction}
Salt, whenever it is present, plays an important role \citep{Salt-IMG} in seismic reflection imaging due to its distinctive acoustic features and usually complex shape. One of the major tasks in seismic imaging and interpretation is then to precisely distinguish salt-bodies from surrounding sediment. In most cases, salt-body possesses a clear boundary and is easy to be identified with human vision. However, seismic data tends to be massive (TB level) and it is not uncommon to have a group of people working for weeks to finish a full-survey salt-body delineating. The clear definition of salt-body and overwhelming amount of data actually makes this challenge a perfect task for deep learning.

Deep learning, which is capable of extracting extremely detailed features from given data, has had a huge impact on the development of image analysis, especially, semantic segmentation. Recently, deep learning found its application in oil and gas industry, such as well log correlation, fault interpretation \citep{ML-method} and rock facies classification\citep{Zeng_rock_facies}. Convolutional Neural Network (CNN), being one of the most powerful 'weapons' in the deep learning arsenal, utilizes numerous convolving/pooling/activation layers to obtain a collection of underlying features from the original image. The effectiveness of CNN in salt-body identification has been shown in a recent study \citep{Y-CNN}, where a proof-of-principle study focusing on factors contributing to the superiority of CNN has been provided.

In this paper, we aim to extend on the work by utilizing the state-of-art CNN with U-Net architecture to fully exploit its potential in regards to salt-body identification. We will first describe the deployed convolutional network structure, and then discuss the adjustments we have made to improve the network training. Finally, we will show the preliminary salt interpretation result and will have some discussions on its possible applications and how to further improve. 

\section{Training and Testing Data}
\begin{figure}[!ht]
  \centering
  \subfloat[Line 1 (inline 4403) seismic image and manually interpreted salt body]{{\includegraphics[scale=0.345]{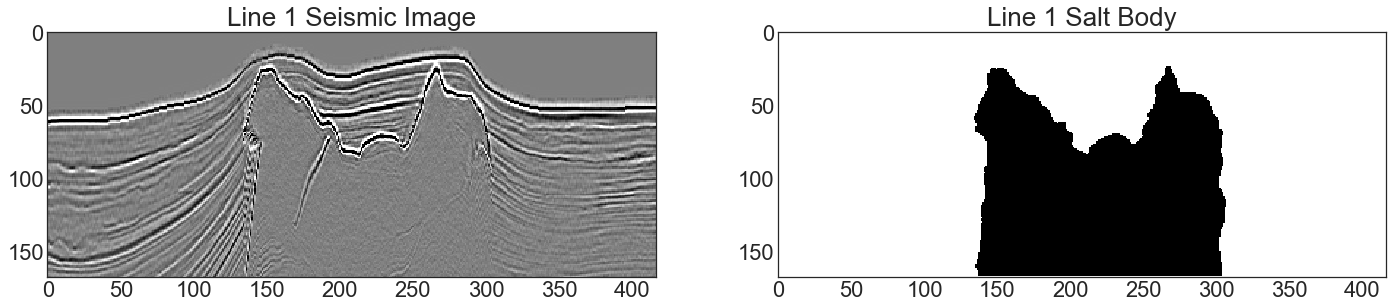}} } \\  
  \subfloat[Line 2 (inline 4499) seismic image and manually interpreted salt body]{{\includegraphics[scale=0.345]{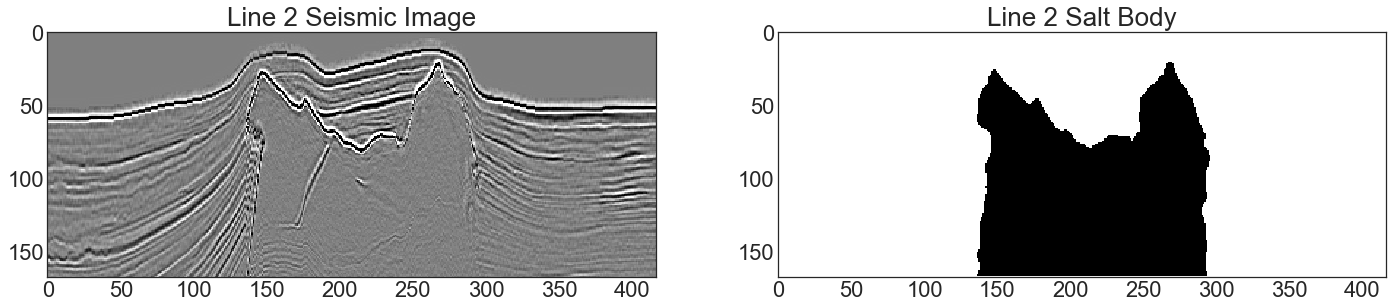}} } \\
  \subfloat[Line 3 (inline 4595) seismic image and manually interpreted salt body]{{\includegraphics[scale=0.345]{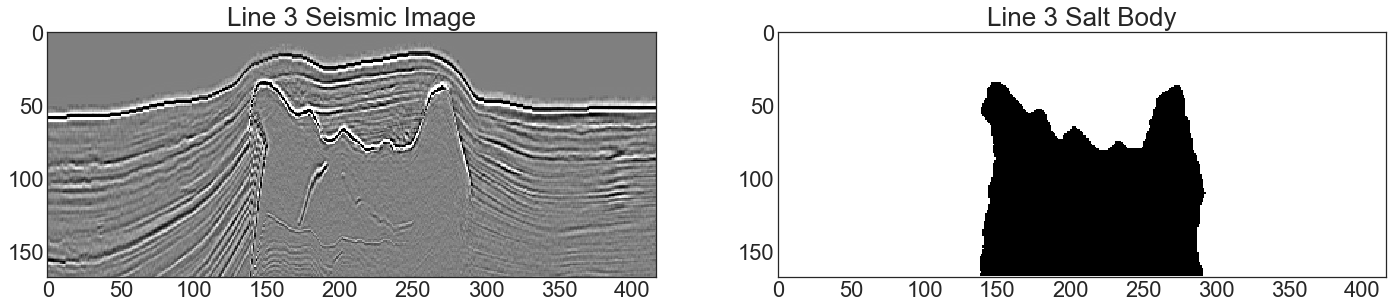}} }
  \caption{Seismic image (left) and the corresponding manually interpreted salt body (right). For salt body interpretation, black color is for salt and white color is for non-salt. Line 1 and 2 are used in training/development; Line 3 is used in testing.}
  \label{fig:3line_qc}
\end{figure}

The dataset we used for this study is the SEG Advanced Modeling (SEAM) Phase 1 data \citep{seam_cite} that emphasizes on deep-water Gulf of Mexico and contains complex salt geometries. 
The salt body on inline number 4403, 4499, and 4595 is manually interpreted by Dr. Haibin Di as in Ref. \citep{Y-CNN}. Both seismic images and corresponding salt body masks are then splitted into small pieces with each piece $101\times101$ pixel in size. For this study, small image pieces from inline 4403 and 4499 are used as training/development set. The small image pieces from inline 4595 are used as testing set. Augmentation including flipping of axes, tilting, rotation and scaling is applied to simulate a larger training sample and to prevent the network from possible overfitting. The seismic images and corresponding salt body labels for these three inlines are shown in Fig.~\ref{fig:3line_qc}.

\section{Network Architecture}
\begin{figure*}[t]
  \centering
  \includegraphics[width=\textwidth]{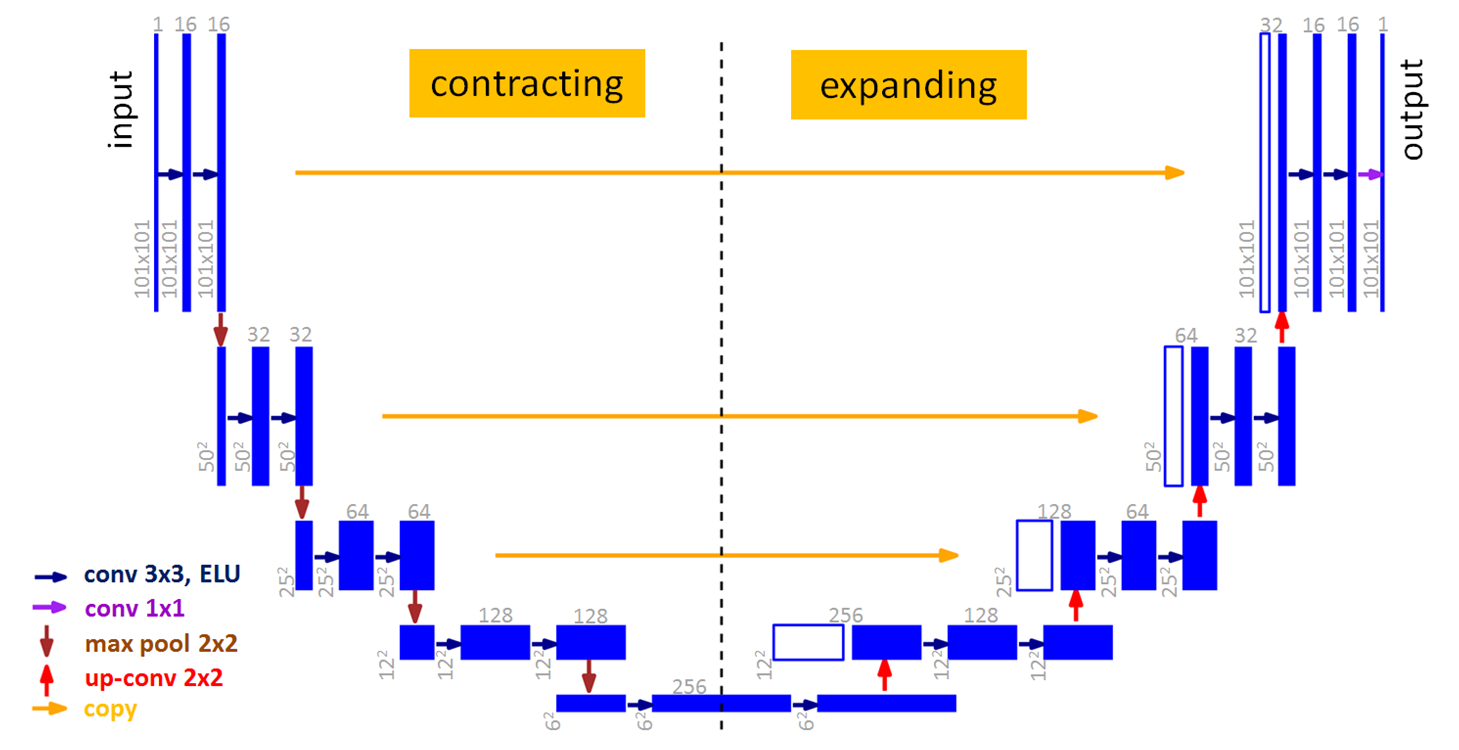}
  \caption{A schematic view of U-Net architecture. Blue box on left corresponds to a multi-channel feature map in contracting path, where the number of channels increases stage by stage. On contrary, blue box on right corresponds to a multi-channel feature map in an expansive path, where the number of channels decreases stage by stage. The horizontal arrow denotes the transfer of residual information from layers in the contracting path to the corresponding layers in the expansive path.}
  \label{fig:U-Net}
\end{figure*}

The U-Net \citep{U-Net} structure, which combines a contracting/down-sampling path to extact context information (what) and a symmetric expanding/up-sampling path to retrieve location information (where), has been used as our baseline architecture. To overcome the degradation problem with deep learning networks, a particular advanced variant of U-Net is adopted in our experiments: the Deep Residual Learning or the ResNet \citep{ResNet}. ResNet is constructed by adding an identity mapping shortcut on top of every few stacked layers. This will result in higher prediction precision due to the fact that it is easier to learn the mapping of the perturbation, which is close to zero, than directly learn from the full input. A schematic view of U-Net architecture is shown in Fig.~\ref{fig:U-Net}.

\section{Evaluation Metric}
In our experimentation, the intersection-over-union (IoU) score, also known as Jaccard index, is used to quantify the performance of CNN salt identification, which is the more widely accepted evaluation matrix in the image segmentation. IoU measures similarity between two or more definitive sample sets. Mathematically, it can be written as the intersection divided by the union of the sample sets. 

\begin{equation}
J(A,B) = \frac{A\cap B}{A\cup B}
\end{equation}

This generates a statistic measure between 0 to 1, with convention of 1 for case 0/0, in a binary segmentation problem. High IoU score (closer to 1) indicates the higher similarity of the two sample sets, thus higher performance of CNN in salt identification. Specifically in this case, higher IoU score means better consistence between the predicted salt pixels and the ground truth salt mask.

\section{Network Training}
In addition to utilizing the state-of-art U-Net + ResNet network structure with conventional configuration for deep neural networks, which includes He initialization \citep{He_initialization} that particularly considers the rectifier nonlinearities, batch normalization and dropout, a few adjustments of the network were implemented to further improve the salt body prediction precision. 

Firstly, a stratified K-folding cross-validation is implemented to split the training data into groups with roughly the same proportions of salt/non-salt images. Compared to regular cross-validation, Stratified cross-validation usually leads to results with smaller bias/variance and therefore better generalization. Secondly, while the standard ReLU activation function may cause some units not being activated at all, we choose to use ELU \citep{ELU} as it acknowledges small negtive values while heavily penalizes large ones. Lastly, we will justify the addition of training epochs in respect to Lov\'{a}sz-Softmax loss. The Jaccard index, or commonly known as the intersection-over-union (IoU) score is boradly accepted as the best evaluation of the precision of image segmentation as demonstrated in \citep{Lovasz}. We also promoted the Lov\'{a}sz-Softmax loss function as an effective surrogate for IoU optimization. Given the fact that the IoU measure is sensitive to hyperparameters such as learning rates and batch sizes, we will train our network initially using cross-entropy loss to locate optimal hyperparameters prior to training against the Lov\'{a}sz-Softmax loss for fine-tuning.


\subsection{Stratified K-Fold}
The $K$-fold cross-validation means splitting the training dataset into $K$-folds, then train on $K-1$ folds and make predictions and evaluations on the remaining 1 fold. The stratified $K$-fold sampling is performed to produce folds that contain a representative ratio of each class and is thus a better way to split training data. As discussed in many machine learning literatures, model averaging is a quite powerful and reliable way to reduce generalization errors. The reason why model averaging can further improve the evaluation score is that different models will unlikely make all the same errors on the test set. Assume each of the $K$ models makes an error $\delta_{i}$ on each event and $\delta_i$ is randomly drawn from a Gaussian distribution with $\mathbb{E}[\delta_i] = 0$, $\mathbb{E}[\delta_i^2] =Var$ and $\mathbb{E}[\delta_i\delta_j] = Cov$. The expected variance of ensemble model will be described by Eqn.\ref{eqn:ensemble}:
\begin{equation}
  \mathbb{E}\left[\frac{1}{K}\left(\sum_i\delta_i\right)^2\right] = \frac{1}{K^2} \mathbb{E}\left[ \sum_{i} \left(\delta_i^2 + \sum_{j\ne i}\delta_i\delta_j \right)\right] = \frac{Var}{K} + \frac{(K-1)\cdot Cov}{K}
  \label{eqn:ensemble}
\end{equation}
In the case where errors are independent, covariance $Cov$ will be 0 and Eqn.\ref{eqn:ensemble} can be simplified as:
\begin{equation}
  \mathbb{E}\left[\frac{1}{K}\left(\sum_i\delta_i\right)^2\right] = \frac{Var}{K}
  \label{eqn:ensemble_ind}
\end{equation}
If errors are perfectly correlated, variance and covariance will be identical and Eqn.\ref{eqn:ensemble} can be simplified as:
\begin{equation}
  \mathbb{E}\left[\frac{1}{K}\left(\sum_i\delta_i\right)^2\right] = Var
  \label{eqn:ensemble_corr_full}
\end{equation}
If errors are partially correlated, the expected variance of ensemble model will be within the two extreme values:
\begin{equation}
  \frac{Var}{K} < \mathbb{E}\left[\frac{1}{K}\left(\sum_i\delta_i\right)^2\right] < Var
  \label{eqn:ensemble_corr_partial}
\end{equation}

For neural networks, the differences in random initialization, random mini-batch selection, and differences in hyperparameter space will tend to make prediction errors from different folds to be partially independent. As demonstrated in Eqn.\ref{eqn:ensemble_corr_partial}, the ensemble of the $K$ models will, on average, perform better than its member models if the errors are not perfectly correlated. 

We have tested stratified $K$-fold training and prediction with $K=5$. The final prediction is the average of the prediction results from 5 models. By comparing with the result not using stratified $K$-fold, we observe a $\sim$ 0.015 improvement in the evaluation score given sufficiently large ($\sim 500$) training epochs.

\subsection{Activation Function}
In neural networks, the activation function of a node is a mathematical function which calculates the output of that node given an input or a set of inputs. The node output will be further used as input for subsequent layers until a desired solution is obtained.

\begin{figure}[!ht]
  \centering
  \subfloat[ReLU activation function]{{\includegraphics[scale=0.6]{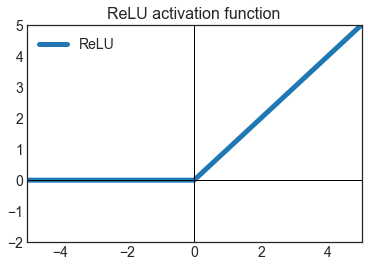}} }%
  \qquad
  \subfloat[ELU activation function]{{\includegraphics[scale=0.6]{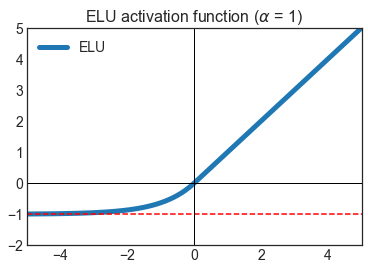}} }%
  \caption{Comparison of the ReLU and the ELU activation functions}
  \label{fig:relu_vs_elu}
\end{figure}

For deep neural networks, one of the known issues is the tricky problem of {\it Vanishing Gradients}. Backpropagation algorithm will compute the gradient of the cost function with regards to each network parameter. In a {\it Gradient Descent} step, these gradients will then be used to update each parameter. However, as the algorithm progresses to the lower layers, the gradients often get smaller and then the parameters are virtually not updated. A poor choice of activation function will make the issue of {\it Vanishing Gradients} even worse. Recent studies \citep{Glorot, Kawaguchi} have demonstrated the {\it Rectified Linear Unit} (ReLU) activation function is better than sigmoid activation function. The ReLU activation function is not only fast to compute but also does not saturate for positive values. Its definition is given by Eqn.\ref{eqn:relu} and is represented in Figure ~\ref{fig:relu_vs_elu} (a).

However, the ReLU activation function still has limitations. One issue of ReLU is that during training, some nodes will only output 0 and will be effectively dead. To overcome this issue of {\it dying ReLUs}, variants of ReLU and other non-ReLU-type activation functions have been proposed. Among those, the {\it Exponential Linear Unit} (ELU) \citep{ELU} activation function is demonstrated to outperform all the ReLU variants. Figure ~\ref{fig:relu_vs_elu} (b) and Eqn.\ref{eqn:elu} gives ELU function's distribution and definition, respectively.

Comparing with the ReLU function, the ELU function has the following characteristics:

\begin{itemize}
\item For $x<0$, the ELU function has non-zero gradient and thus fix the issue of dying nodes observed in the ReLU function.
\item The ELU function is smooth everywhere even at $x=0$. This may help to speed up the {\it Gradient Descent} calculation.
\item The ELU function generally converges faster than the ReLU-type functions during training.
\item Due to its exponential nature, the ELU function is generally slower in computation than the ReLU function and its variants. 
\end{itemize}

In this seismic salt interpretation experiment, the ReLU function is initially used for the first 500 training epochs for better runtime performance. For later trainings, the ELU activation function is used instead. Comparing with using the ReLU function only, we found the switch from ReLU to ELU can improve the IoU score by $\sim$ 0.002.

\begin{equation}
  \text{ReLU}(x)=\begin{cases}
             0 & \text{if } x<0 \\
             x & \text{if } x\ge 0
          \end{cases}
  \label{eqn:relu}
\end{equation}

\begin{equation}
  \text{ELU}_{\alpha}(x)=\begin{cases}
                  \alpha\cdot(e^x -1) & \text{if } x<0 \\
                  x & \text{if } x\ge 0
                  \end{cases}
  \label{eqn:elu}
\end{equation}

\subsection{Loss Function}
Conventional choice of loss function for neural network is to optimize the cross-entropy\citep{Xentropy} based on logistic regression. Complementary to the IoU score (or Jaccard index), Jaccard loss, obtained by subtracting the IoU score from 1, can be used to measure dissimilarity between sample sets and has been shown to be a better evaluation for image segmentation tasks \citep{Lovasz}. Therefore, maximizing IoU score is equivalent to minimizing Jaccard loss. The latter can be constructed as the loss function for the image segmentation. 

It is well known that convex surrogate loss functions are essential to the practical application of empirical loss minimization\citep{convex}. The issue of Jaccard loss is submodular. In other words, the Jaccard loss doesn't converge very well. To this end, Ref.\citep{Lovasz} applies the Lov\'{a}sz hinge\citep{convex} as a good surrogate to Jaccard loss, and yields a consistent improvement in capturing the absolute minimum of the Jaccard loss. The test datasets with this Lov\'{a}sz hinge Jaccard loss in the same reference shows benefits like better recovery of small objects and filling more gaps inside a large continum. Lov\'{a}sz hinge Jaccard loss function can be generalized to multi-class classification problem as Lov\'{a}sz-Softmax loss function by considering the class-averaged IoU metric (mIoU):
\begin{equation}
loss(f) = \frac{1}{|C|}\sum_{c\subset C}{\overline{\bigtriangleup_{J_c}}}(m(c))
\end{equation}
where $\bigtriangleup_{J_c}$ is the loss surrogate, $m(c)$ corresponds to the vector of pixel identification errors, and $c$ is one subclass of multiple classes of $C$. Comparing with the usage of conventional cross-entropy loss, we find the usage of the Lov\'{a}sz-Softmax loss can improve the prediction accuracy by $\sim$0.002.

Though Lov\'{a}sz-Softmax loss function used in the U-Net has been shown \citep{lovas_app} to alleviate obstacles in computing vision classification task, like small amount of data, incomplete or mislabeling, and highly imbalanced classes, it is however, as discussed in Ref.\citep{Lovasz}, sensitive to hyper-parameters such as learning rates and batch sizes. In our trainings, cross-entropy loss is thus first used to make hyper-parameters converge toward optimal values. Later, in fine-tuning stage, the Lov\'{a}sz-Softmax loss is used in order to achieve better prediction accuracy. 

\subsection{Learning Rate Scheduling}
\begin{figure}[!ht]
  \centering
  \includegraphics[scale=0.7]{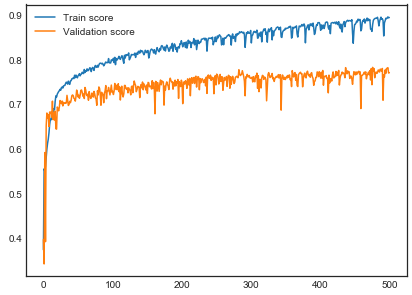}
  \caption{Train score and validation score for 500 training epochs using cross-entropy loss and learning rate $\eta_1=0.001$}
  \label{fig:loss1}
\end{figure}

\begin{figure}[!ht]
  \centering
  \includegraphics[scale=0.7]{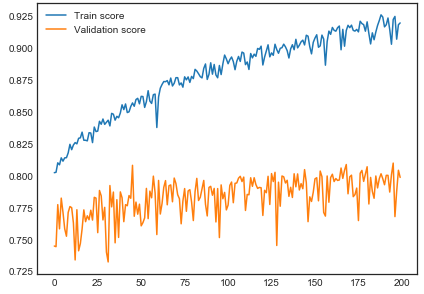}
  \caption{Train score and validation score for the first 200 training epochs using Lov\'{a}sz loss and learning rate $\eta_2=0.0005$}
  \label{fig:loss2}
\end{figure}

\begin{figure}[!ht]
  \centering
  \includegraphics[scale=0.7]{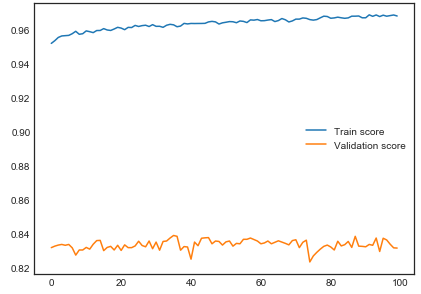}
  \caption{Train score and validation score for 100 training epochs using Lov\'{a}sz loss and learning rate $\eta_3=0.0001$}
  \label{fig:loss3}
\end{figure}

The ideal learning rate has the advantage of learning quickly and converge to optimal solution. However, finding a good learning rate can be tricky. In this salt body interpretation experiment, we adopted a simple strategy called {\it predetermined piecewise constant learning rate}. The initial learning rate is set to $\eta_1 = 0.001$ for the first 500 epochs, then to $\eta_2=0.0005$ for the next 400 epochs and to $\eta_3=0.0001$ for the next 100 epochs. For each learning rate scenario, the best trained models are saved and are later reloaded for a different learning rate scenario. Table~\ref{tab:train_para} summarizes key training parameters and algorithms for different training epochs. Training and validation score curves as a function of training epoch number are shown in Figures ~\ref{fig:loss1}, ~\ref{fig:loss2} and ~\ref{fig:loss3}. A general trend of improved IoU score with more training epochs is observed. 

\begin{table*}[t]
\centering
{\begin{tabular}{@{}cccccc@{}} \toprule
\hline
Training Epoch & Learning Rate & Loss Function & Algorithm \\
 \hline
1-500 & 0.001 & Cross Entropy & Adam     \\
501-900 & 0.0005 & Lov\'{a}sz & Adam \\
901-1000 & 0.0001 & Lov\'{a}sz & Adam \\
 \hline
\end{tabular} }
\caption{Learning rate, loss function and algorithm used for different training epochs.}
\label{tab:train_para}
\end{table*}

\section{Results}
\begin{figure*}[t]
  \centering
  \includegraphics[width=\textwidth]{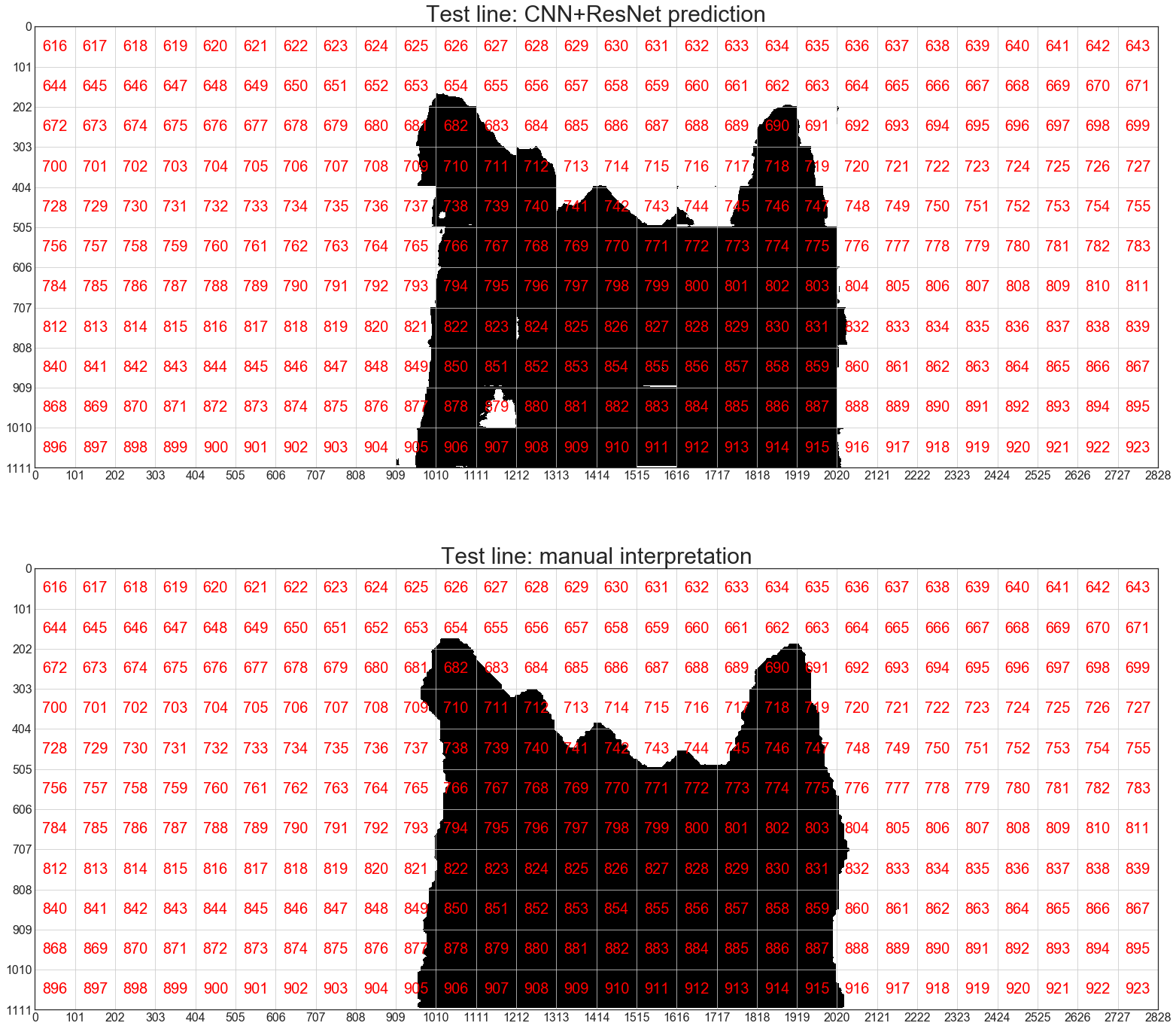}
  \caption{Comparison of manual interpretation and CNN-based prediction with 1000 training epochs.}
  \label{fig:true_vs_pred}
\end{figure*}

To assess the quality of the automatic seismic interpretation, the best trained stratified 5-fold models are averaged and applied to the testing inline 4595. The comparison of its salt body prediction and a manual interpretation is shown in Fig.~\ref{fig:true_vs_pred}.

Overall, the general salt body shape predicted by CNNs agrees quite well with manual interpretation. This demonstrates the great potential in computer-aided seismic interpretation using the CNNs. We also notice that details at salt boundaries, though quite similar as manual interpretation, are not identical in some local places. What's more, areas with weak seismic reflection, for example block 879 in Fig.~\ref{fig:true_vs_pred}, may still suffer from mis-classification. 



\section{Discussion}
For the relatively complex CNN architecture we have adopted and given the inputs we already have, we find the major improvements are mostly coming from more training epochs. In comparison, though the IoU improvements introduced by using ELU activation function and by using Lov\'{a}sz loss function are relatively small ($\sim$0.002 improvement for each), it is however robust and non-negligible. Ensemble method using stratified $K$-fold can further improve the test score by $\sim$0.015.

We would expect more training samples, the usage of pre-trained netwroks (i.e. ResNet34), and the implementation of snapshot ensemble would further improve the prediction accuracy. To speed up the training process and make the models converge faster, strategy of cyclic learning rate could be adopted to further improve. Another possible way to further improve, is by constructing edge detection related features as additional channel for the CNN. 

At current stage, the CNN-based salt body interpretation is still not perfect and cannot replace manual interpretation. However, the implementation of deep CNNs on seismic data for salt body identification is still promising. For example, the rough salt body derived from CNN can be used as input for {\it Full Waveform Inversion} \citep{FWI} without the need to manually interpret a salt body. Another possible application of this CNN-based method would be the real-time feature segmentation.

\section{Conclusion}
We have implemented a deep CNN-based architecture for automatic seismic data salt body interpretation. The network structure is based on U-net plus ResNet using the Adam algorithm. A few improvements, including ELU activation function, Lov\'{a}sz cost function, stratified $K$-fold cross validation and model averaging, have been made to further improve the salt body prediction accuracy. We have demonstrated that salt body can be successfully delineated using seismic data alone in an automatic manner and the prediction result is confirmed to agree well with a manually interpreted salt body. 

\section{ACKNOWLEDGMENTS}
The authors would like to thank SEG Advanced Modeling Corporation for providing SEAM data available for public usage, and Dr. Haibin Di for providing the training and the testing datasets used for this R\&D. The neural network algorithm is implemented using the open-source Tensorflow package developed by Google.

\bibliographystyle{seg} 

\end{document}